\documentclass[preprint,twocolumn]{aastex63}
\usepackage[utf8]{inputenc}

\usepackage{xcolor}
\usepackage{multirow}

\newcommand{\scone}{{\small SCONE}}

\graphicspath{{.}}

\begin{document}
\bibliographystyle{aasjournal}

\title{\scone: Supernova Classification with a Convolutional Neural Network}

\author[0000-0003-1899-9791]{Helen Qu}
\affiliation{Department of Physics and Astronomy, University of Pennsylvania, Philadelphia, PA 19104, USA}
\author[0000-0003-2764-7093]{Masao Sako}
\affiliation{Department of Physics and Astronomy, University of Pennsylvania, Philadelphia, PA 19104, USA}
\author[0000-0001-8211-8608]{Anais M{\"o}ller}
\affiliation{LPC, Université Clermont Auvergne, CNRS/IN2P3, F-63000 Clermont-Ferrand, France}
\author[0000-0003-4480-0096]{Cyrille Doux}
\affiliation{Department of Physics and Astronomy, University of Pennsylvania, Philadelphia, PA 19104, USA}

\correspondingauthor{Helen Qu}
\email{helenqu@sas.upenn.edu}

\begin{abstract}
    We present a novel method of classifying Type Ia supernovae using convolutional neural networks, a neural network framework typically used for image recognition. Our model is trained on photometric information only, eliminating the need for accurate redshift data. Photometric data is pre-processed via 2D Gaussian process regression into two-dimensional images created from flux values at each location in wavelength-time space. These ``flux heatmaps" of each supernova detection, along with ``uncertainty heatmaps" of the Gaussian process uncertainty, constitute the dataset for our model. This preprocessing step not only smooths over irregular sampling rates between filters but also allows \scone\ to be independent of the filter set on which it was trained. Our model has achieved impressive performance without redshift on the in-distribution SNIa classification problem: $99.73 \pm 0.26$\% test accuracy with no over/underfitting on a subset of supernovae from PLAsTiCC's unblinded test dataset. We have also achieved $98.18 \pm 0.3$\% test accuracy performing 6-way classification of supernovae by type. The out-of-distribution performance does not fully match the in-distribution results, suggesting that the detailed characteristics of the training sample in comparison to the test sample have a big impact on the performance.  We discuss the implication and directions for future work. All of the data processing and model code developed for this paper can be found in the \href{https://github.com/helenqu/scone\ }{\scone\ software package} located at github.com/helenqu/scone \citep{helen_qu_2021_4660288}.
\end{abstract}


\section{introduction}
The discovery of the accelerating expansion of the universe \citep{perlmutter,riess} has led to an era of sky surveys designed to probe the nature of dark energy. Type Ia supernovae (SNe Ia) have been instrumental to this effort due to their standard brightness and light curve profiles. Building a robust dataset of SNe Ia across a wide range of redshifts will allow for the construction of an accurate Hubble diagram that will enrich our understanding of the expansion history of the universe as well as place constraints on the dark energy equation of state.

Modern-scale sky surveys, including SDSS, Pan-STARRS, and the Dark Energy Survey, have identified thousands of supernovae throughout their operational lifetimes \citep{sdss,panstarrs,des}. However, it has been logistically challenging to follow up on most of these detections spectroscopically. The result is a low number of spectroscopically confirmed SNe Ia and a large photometric dataset of SNe Ia candidates. The upcoming Rubin Observatory Legacy Survey of Space and Time (LSST) is projected to discover $10^7$ supernovae \citep{lsstsciencecollaboration2009lsst}, with millions of transient alerts each observing night. As spectroscopic resources are not expected to scale with the size of these surveys, the ratio of spectroscopically confirmed SNe to total detections will continue to shrink. With only photometric data, distinguishing between SNe Ia and other types can be difficult. A reliable photometric SNe Ia classification algorithm will allow us to tap into the vast potential of the photometric dataset and pave the way for confident classification and analysis of the ever-growing library of transients from current and future sky surveys.

Significant progress has been made in the past decade in the development of such an algorithm. Most approaches involve lightcurve template matching \citep{sako}, or feature extraction paired with either sequential cuts \citep{bazin,campbell} or machine learning algorithms \citep{Moller_2016,Lochner_2016,Dai_2018,Boone_2019}. Most recently, the spotlight has been on deep learning techniques since it has been shown that classification based on handcrafted features is not only more time-intensive for the researcher but is outperformed by neural networks trained on raw data \citep{Charnock_2017,moss,Kimura_2017}. Since then, many neural network architectures have been explored for SN photometric classification, such as PELICAN's CNN architecture \citep{pasquet} and SuperNNova's deep recurrent network \citep{supernnova}.

Several photometric classification competitions have been hosted, including the Supernova Photometric Classification Challenge (SPCC) \citep{spcc} and the Photometric LSST Astronomical Time Series Classification Challenge (PLAsTiCC) \citep{plasticc}. These have not only resulted in the development of new techniques, such as PSNID \citep{sako} and Avocado \citep{Boone_2019}, but have also provided representative datasets available to researchers during and after the competition, such as the PLAsTiCC unblinded dataset used in this paper.

In this paper we present \scone, a novel application of deep learning to the photometric classification problem. \scone is a convolutional neural network (CNN), an architecture prized in the deep learning community for its state-of-the-art image recognition capabilities \citep{lecun1989backpropagation,lecun1998gradient,alexnet,zeiler2014visualizing,vgg}. Our model requires raw photometric data only, precluding the necessity for accurate redshift approximations. The dataset is preprocessed using a lightcurve modeling technique via Gaussian processes described in \citet{Boone_2019}, which alleviates the issue of irregular sampling between filters but also allows the CNN to learn from information in all filters simultaneously. The model also has relatively low computational and dataset size requirements without compromising on performance -- 400 epochs of training on our $\sim 10^4$ dataset requires around 15 minutes on a GPU.

We will introduce the datasets used to train and evaluate \scone\, its computational requirements, as well as the algorithm itself in Section 2. Section 3 will focus on the performance of \scone\ on both binary and categorical classification, and Section 4 presents an analysis of misclassified lightcurves and heatmaps for both modes of classification.
\\\\\\
\section{methods}
\subsection{Datasets}
The PLAsTiCC training and test datasets were originally created for the 2018 Photometric LSST Astronomical Time Series Classification Challenge.

The PLAsTiCC training set includes $\sim 8000$ simulated observations of low-redshift, bright astronomical sources, representing objects that are good candidates for spectroscopic follow-up observation. This dataset will be referred to in future sections as the ``spectroscopic dataset". We use this dataset to evaluate the out-of-distribution performance of \scone\ in section 3.1.2.

The PLAsTiCC test set consists of 453 million simulated observations of 3.5 million transient and variable sources, representing 3 years of expected LSST output \citep{plasticc_sims}. The objects in this dataset are generally fainter, higher redshift, and do not have associated spectroscopy. Note that most of the results presented in this paper are produced from this dataset alone and will be referred to as ``the dataset", ``the main dataset", or ``the PLAsTiCC dataset" in future sections.

All observations in both datasets were made in LSST's $ugrizY$ bands and realistic observing conditions were simulated using the LSST Operations Simulator \citep{delgado}. While PLAsTiCC includes data from many other transient sources, we are using only the supernovae in the datasets. We selected all of the type II, Iax, Ibc, Ia-91bg, Ia, and SLSN-1 sources (corresponding to \texttt{true\_target} values of 42, 52, 62, 67, 90, and 95) from this dataset and chose only well-sampled lightcurves by restricting ourselves to observations simulating LSST's deep drilling fields (\texttt{ddf=1}). \texttt{peak\_mjd} values, the modified Julian date of peak flux for each object, were calculated for both the main dataset and the spectroscopic dataset by taking the signal-to-noise weighted average of all observation dates. \texttt{peak\_mjd} is referred to as $t_{\rm peak}$ in future sections. The total source count for the spectroscopic set is 4,556, and the total source count for the main dataset is 32,087. A detailed breakdown by type is provided in Table~\ref{tbl:fulldata}.

The categorical dataset was created using SNANA \citep{snana} with the same models and redshift distribution as the main dataset in order to perform categorical classification with balanced classes. 2,000 examples of each type were randomly selected to constitute a class-balanced dataset of 12,000 examples.

\begin{table}
    \centering
    \caption{Makeup of the PLAsTiCC dataset by type.}
    \begin{tabular}{c c c}
        \hline
        SN type & \multicolumn{2}{c}{number of sources}\\
        \hline
         & spectroscopic & main \\
        \hline
        SNIa & 2,313 & 12,640 \\
        SNII & 1,193 & 15,986 \\
        SNIbc & 484 & 2,194 \\
        SNIa-91bg & 208 & 362 \\
        SNIax & 183 & 807 \\
        SLSN-1 & 175 & 98 \\
        \hline
    \end{tabular}
    \label{tbl:fulldata}
\end{table}

\subsection{Quality Cuts}
In order to ensure that the model is learning only from high-quality information, we have instituted some additional quality-based cuts on all datasets. These cuts are based on lightcurve quality, so all metrics are defined for a single source. The metrics evaluated for these cuts are as follows:
\begin{itemize}
    \item \textbf{number of detection datapoints ($n_{\rm detected}$)}: number of observations where the source was detected. We chose a detection threshold of S/N $>$ 5, based on Fig. 8 of \citet{kessler_2015}
    \item \textbf{cumulative signal-to-noise ratio ($CSNR$)}: cumulative S/N for all points in the lightcurve
    $$CSNR = \sqrt{\sum{\frac{f^2}{\sigma_f^2}}}$$
    \item \textbf{duration}: timespan of detection datapoints
    $$t_{\rm active} = t_{\rm last} - t_{\rm first}$$
\end{itemize}
where $f$ represents the flux measurements from all observations of a given source, $\sigma_f$  represents the corresponding uncertainties, $t$ represents the timestamps of all observations of a given source, $t_{\rm first}$ is the time of initial detection, and $t_{\rm last}$ is the time of final detection. Our established quality thresholds require that:
\begin{itemize}
    \item $n_{\rm detected} \geq 5$
    \item $CSNR > 10$
    \item $t_{\rm active} \geq 30$ days
\end{itemize}
for lightcurve points in the range $t_{\rm peak}-50 \leq t \leq t_{\rm peak}+130$.
1,150 out of 4,556 sources passed these cuts in the spectroscopic dataset and 12,611 out of 32,087 sources passed these cuts in the main dataset. The makeup of these datasets is detailed in Table \ref{tbl:cutdata}. The categorical dataset was created from sources that already passed the cuts, so the makeup is unchanged.

\subsection{Class Balancing}

Maintaining an equal number of examples of each class, or a balanced class distribution, is important for machine learning datasets. Balanced datasets allow for an intuitive interpretation of the accuracy metric as well as provide ample examples of each class for the machine learning model to learn from.

As shown in Table \ref{tbl:cutdata}, the natural distribution of the spectroscopic dataset is more abundant in Ia sources than non-Ia sources. Thus, all non-Ia sources were retained in the class balancing process for binary classification and an equivalent number of Ia sources were randomly chosen. SNIax and SNIa-91bg sources were labeled as non-Ia sources for binary classification. The class-balanced spectroscopic dataset has 496 sources of each class for a total of 992 sources.

In contrast, the natural distribution of the main dataset is more abundant in non-Ia sources than Ia sources. Thus, all Ia sources were retained in the class balancing process for binary classification and an equivalent number of non-Ia sources were randomly chosen. The random selection process does not necessarily preserve the original distribution of non-Ia types. The class-balanced dataset has 6,128 sources of each class for a total of 12,256 sources.

The categorical dataset of 2,000 sources for each of the 6 types was created explicitly for the purpose of retaining balanced classes in categorical classification, as mentioned in Section 2.1.

\begin{table}
    \centering
    \caption{Makeup of the PLAsTiCC dataset by type after applying quality cuts.}
    \label{tbl:cutdata}
    \begin{tabular}{c c c}
        \hline
        SN type & \multicolumn{2}{c}{number of sources}\\
        \hline
         & spectroscopic & main \\
        \hline
        SNIa & 654 & 6,128 \\
        SNII & 262 & 5,252 \\
        SNIbc & 97 & 779 \\
        SNIa-91bg & 41 & 113 \\
        SNIax & 59 & 281 \\
        SLSN-1 & 37 & 58 \\
        \hline
    \end{tabular}
\end{table}

All datasets were split by class into 80\% training, 10\% validation, and 10\% testing. Splitting by class ensures balanced classes in each of the training, validation, and test sets.

\begin{figure}
    \figurenum{1}
    \label{fig:lightcurves}
    \includegraphics[scale=0.13,trim={6cm 33cm 0 0}]{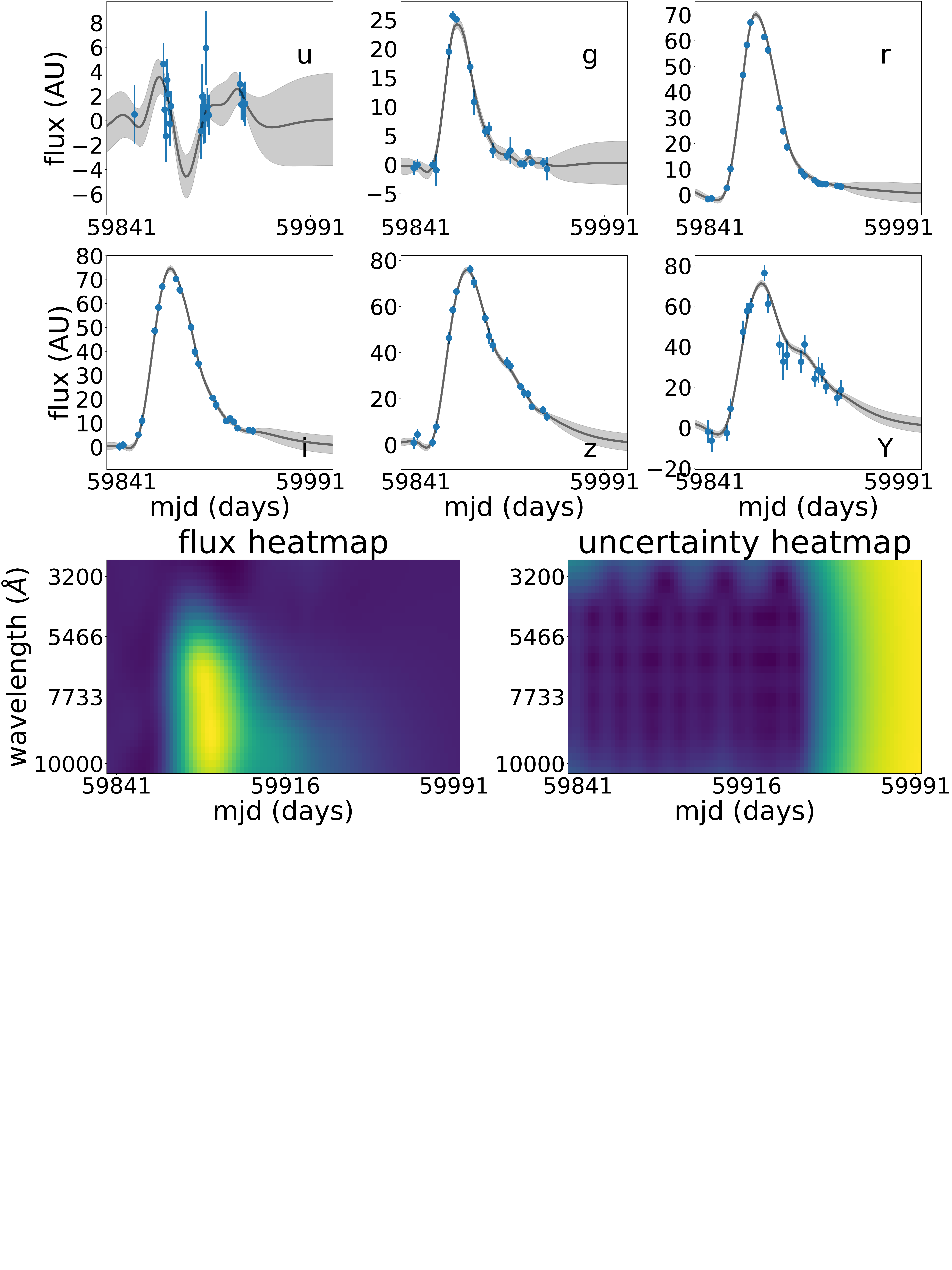}
    \centering
    \caption{Raw $ugrizY$ lightcurve data with the Gaussian process model at corresponding wavelengths and resulting heatmap and error heatmap for a type Ia SN. The shaded regions in the Gaussian process plots represent the Gaussian process error.}
\end{figure}

\subsection{Heatmap Creation}
Prior to training, we preprocess our lightcurve data into heatmap form. First, all observations are labeled with the central wavelength of the observing filter according to Table \ref{tbl:filters}, which was calculated from the filter functions used by \citet{plasticc}. We then use the approach described by \citet{Boone_2019} to apply 2-dimensional Gaussian process regression to the raw lightcurve data to model the event in the wavelength ($\lambda$) and time ($t$) dimensions. We use the Matern 32 kernel with a fixed 6000~\AA\  characteristic length scale in $\lambda$ and fit for the length scale in $t$. Once the Gaussian process model has been trained, we obtain its predictions on a $\lambda,t$ grid and call this our ``heatmap". Our choice for the $\lambda,t$ grid was $t_{\rm peak} - 50 \leq t \leq t_{\rm peak} + 130$ with a 1-day interval and $3000 < \lambda < 10,100$~\AA\ with a $221.875$~\AA\ interval. The significance of this choice is explored further in Section 3.

\begin{figure*}
    \figurenum{2}
    \label{fig:heatmaps}
    \includegraphics[scale=0.35,trim={5cm 1cm 0 0}]{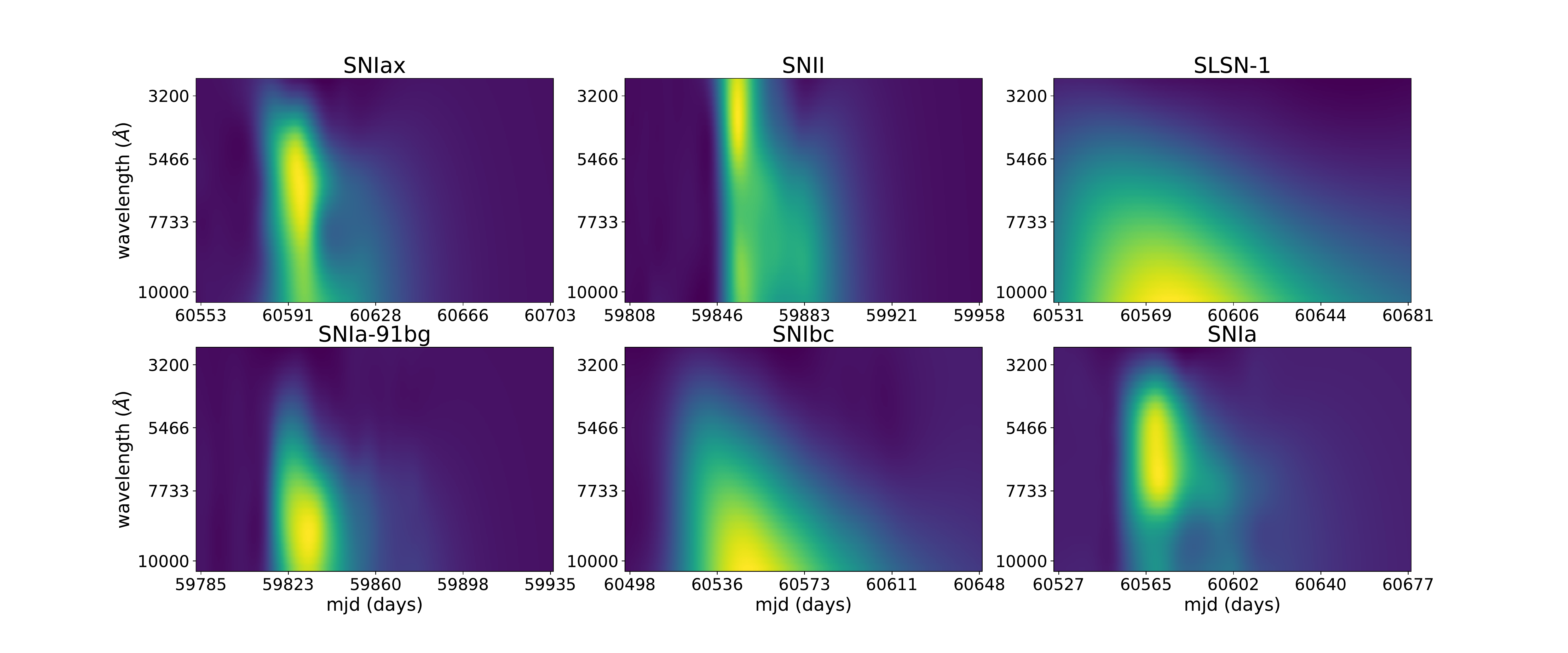}
    \caption{Example flux heatmaps for each supernova type.}
\end{figure*}

\begin{table}
    \centering
    \caption{Central wavelength of each filter}
    \label{tbl:filters}
    \begin{tabular}{c c}
        \hline
        Filter & Central Wavelength (\AA) \\
        \hline
        u & 3670.69 \\
        g & 4826.85 \\
        r & 6223.24 \\
        i & 7545.98 \\
        z & 8590.90 \\
        Y & 9710.28 \\
        \hline
    \end{tabular}
\end{table}

The result of this is a $n_{\lambda} \times n_t$ image-like matrix, where $n_{\lambda}$ and $n_t$ are the lengths of the wavelengths array and times array, respectively, given to the Gaussian process. We also take into account the uncertainties on the Gaussian process predictions at each time and wavelength, producing a second image-like matrix. We stack these two matrices depthwise and divide by the maximum flux value to constrain all entries to [0,1]. This matrix is our input to the convolutional neural network.

Figure~\ref{fig:lightcurves} shows the raw lightcurve data in the $t_{\rm peak} - 50 \leq t \leq t_{\rm peak} + 130$ range for each filter in blue, the Gaussian process model in gray, and the resulting flux and uncertainty heatmaps at the bottom. Figure~\ref{fig:heatmaps} shows a representative example of a heatmap for each supernova type.

\subsection{Convolutional Neural Networks}
Convolutional neural networks (CNNs) are a type of artificial neural network that makes use of the convolution operation to learn local, small-scale structures in an image. This is paired with an averaging or sub-sampling layer, often called a \textit{pooling layer}, that reduces the resolution of the image and allows the subsequent convolutional layers to learn hierarchically more complex and less localized structures.

In a \textit{convolutional layer}, each unit receives input from only a small neighborhood of the input image. This use of a restricted receptive field, or \textit{kernel size}, resembles the neural architecture of the animal visual cortex and allows for extraction of local, elementary features such as edges, endpoints, or corners. All units, each corresponding to a different small neighborhood of the image, share the same set of learned weights. This allows them to detect the presence of the same feature in each neighborhood of the image. Each convolutional layer often has several layers of these units, each of which is called a \textit{filter} and extracts a different feature. The output of a convolutional layer is called a \textit{feature map}.

Pooling layers reduce the local precision of a detected feature by sub-sampling the each unit's receptive field, often $2 \times 2$ pixels, based on some rule. Average-pooling, for example, extracts the average of the 4 pixels, and max-pooling extracts the maximum value. Assuming no overlap in the receptive fields, the spatial dimensions of the resulting feature maps will be reduced by half.

Dropout \citep{dropout} is a commonly used regularization technique in fully connected layers. A \textit{dropout layer} chooses a random user-defined percentage of the input weights to set to zero, improving the robustness of the learning process.

A convolutional neural network typically consists of alternating convolutional layers and pooling layers, followed by a series of fully-connected layers that learn a mapping between the result of the convolutions and the desired output.

\subsection{\scone\ Architecture}

\begin{figure}[h!]
    \figurenum{3}
    \label{fig:architecture}
    \vspace*{0.2cm}
    \includegraphics[scale=0.8,trim={2cm 0 2cm 0}]{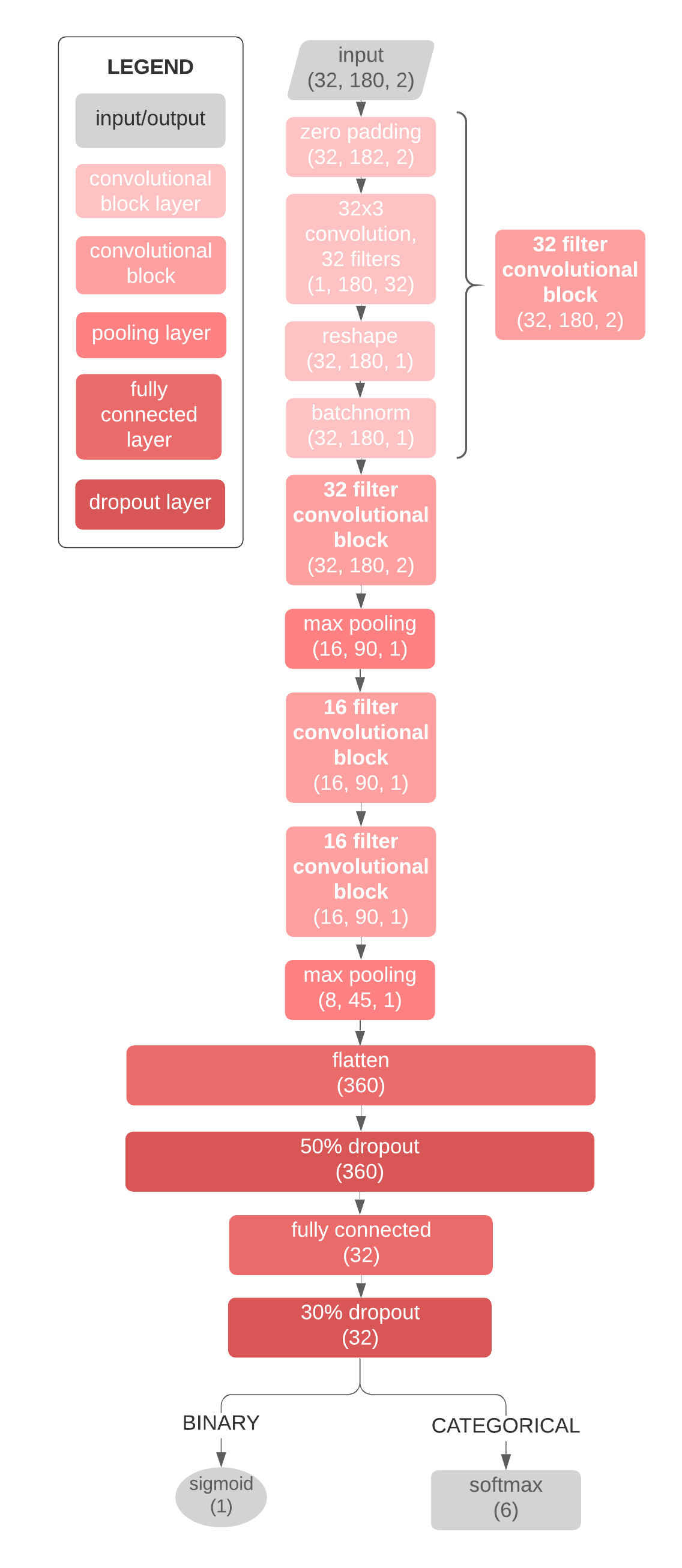}
    \centering
    \caption{\scone\ architecture for binary and categorical classification.}
\end{figure}

The relatively simple architecture of \scone, shown in Figure~\ref{fig:architecture}, allows for a minimal number of trainable parameters, speeding up the training process significantly without compromising on performance. It has a total of 22,606 trainable parameters for categorical classification and 22,441 trainable parameters for binary classification when trained on heatmaps of size $32 \times 180 \times 2$ $(h \times w \times d)$.

As mentioned in Section 2.4, each heatmap is divided by its maximum flux value for normalization. After receiving the normalized heatmap as input, the network pads the heatmap with a column of zeros on both sides, bringing the heatmap size to $32 \times 182 \times 2$. Then, a convolutional layer is applied with $h$ filters and a kernel size of $h \times 3$, which in this case is 32 filters and a $32 \times 3$ kernel, resulting in a feature map of size $1 \times 180 \times 32$. We reshape this feature map to be $32 \times 180 \times 1$, apply batch normalization, and repeat the above process one more time. We have now processed our heatmap through two ``convolutional blocks" with an output feature map of size $32 \times 180 \times 1$.

We apply $2 \times 2$ max pooling to our output, reducing its dimensions to $16 \times 90 \times 1$, and pass it through two more convolutional blocks, but this time $h=16$.

We pass our output through a final $2 \times 2$ max pooling layer, resulting in a $8 \times 45 \times 1$ feature map. This is subsequently flattened into a 360-element array and passed through a 50\% dropout layer. A 32-unit fully connected layer followed by a 30\% dropout layer feeds into the final layer. For binary classification, this is a node with a sigmoid activation that returns the model's predicted Ia probability. For categorical classification, the final layer contains 6 nodes with softmax activations that return the respective probabilities of each of the 6 SN types. Both of these versions of \scone\ are shown in  Figure~\ref{fig:architecture}.

The model is trained with the binary crossentropy loss function for binary classification and the sparse categorical crossentropy loss function for categorical classification. Both classification modes use the Adam optimizer \citep{kingma2017adam} at a constant 1e-3 learning rate for 400 epochs.

\subsection{Evaluation Metrics}

\textit{Accuracy} is defined as the number of correct predictions divided by the number of total predictions. We also evaluated the model on a number of other performance metrics: purity, efficiency, and AUC. 

\textit{Purity} and \textit{efficiency} are defined as: $$\textrm{purity=}\frac{\textrm{TP}}{\textrm{TP+FP}};\; \textrm{efficiency=}\frac{\textrm{TP}}{\textrm{TP+FN}}$$
where TP is true positive, FP is false positive, and FN is false negative.

\textit{AUC}, or \textit{area under the receiver operating characteristic (ROC) curve}, is a common metric used to evaluate binary classifiers. The ROC curve is created by plotting the false positive rate against the true positive rate at various discrimination thresholds, showing the sensitivity of the classifier to the chosen threshold. A perfect classifier would score a 1.0 AUC value while a random classifier would score a 0.5.

\subsection{Computational Requirements}

Due to the minimal number of trainable parameters and dataset size, the time and hardware requirements for training and evaluating with \scone\  are relatively low. The first training epoch on one NVIDIA V100 Volta GPU takes approximately 2 seconds, and subsequent training epochs take approximately 1 second each with TensorFlow's dataset caching. Training epochs on one Haswell node (with Intel Xeon Processor E5-2698 v3), which has 32 cores, take approximately 26 seconds each.

\section{results}

\subsection{Binary Classification}
Our model achieved $99.93 \pm 0.06$\% training accuracy, $99.71 \pm 0.2$\% validation accuracy, and $99.73 \pm 0.26$\% test accuracy on the Ia vs. non-Ia binary classification problem performed on the class-balanced dataset of 12,256 sources. 

\begin{figure}
    \figurenum{4}
    \label{fig:binary-confusion}
    \includegraphics[scale=0.51]{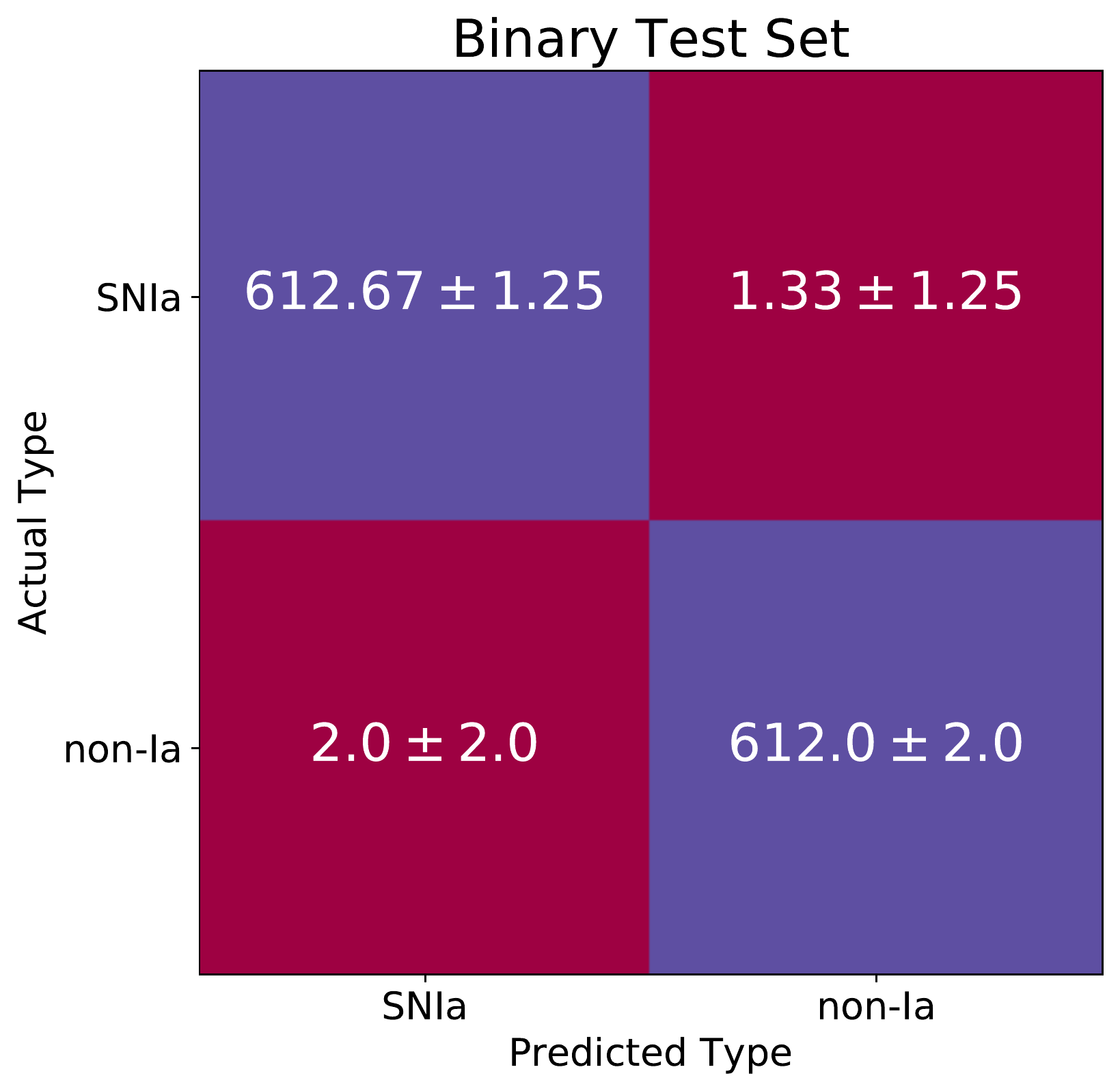}
    \centering
    \caption{Confusion matrix showing average and standard deviation over five runs for binary classification on the test set.}
\end{figure}

\begin{figure}
    \figurenum{5}
    \label{fig:roc}
    \includegraphics[scale=0.35]{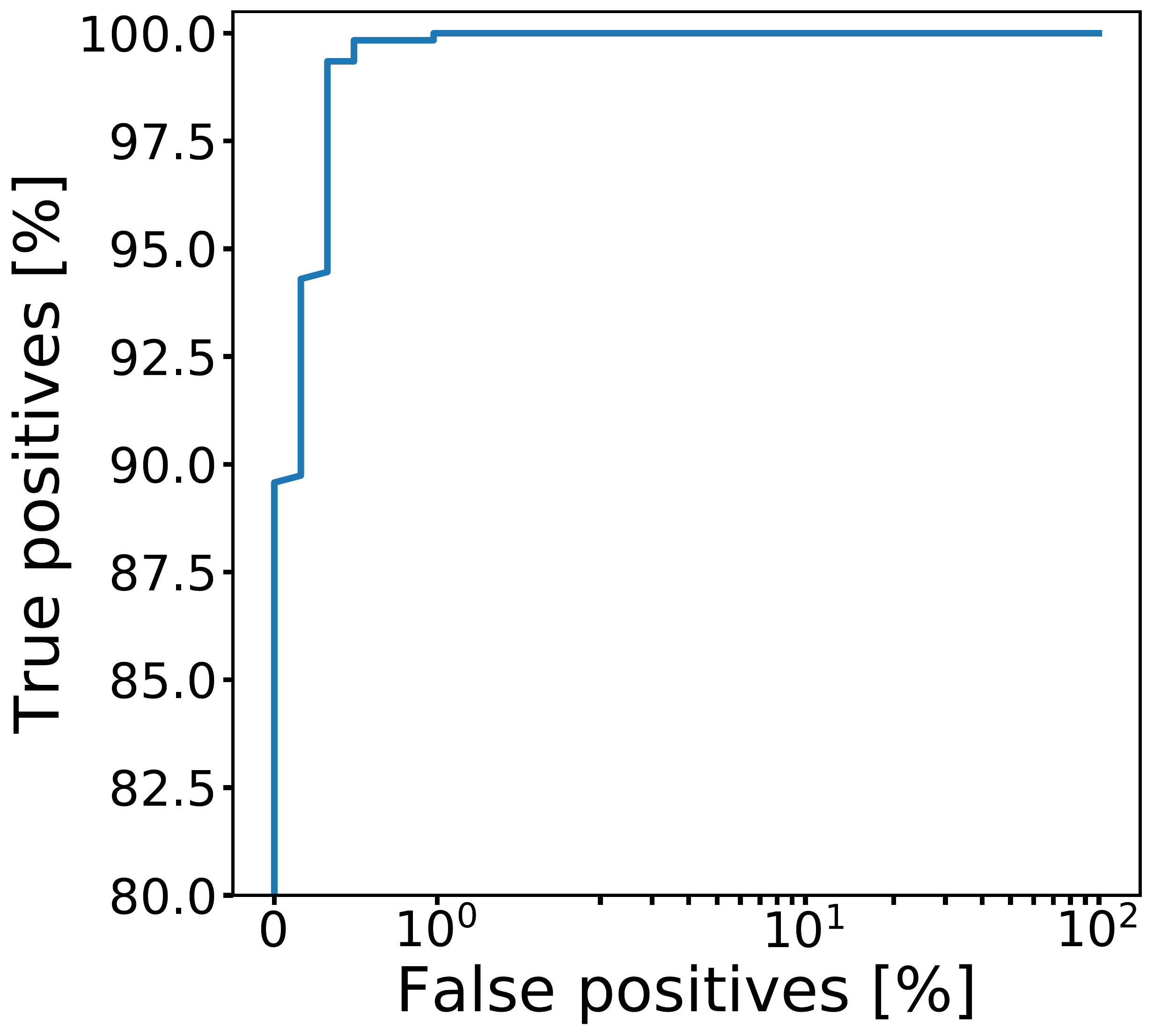}
    \centering
    \caption{Semilog plot of the ROC curve for binary classification on the test set.}
\end{figure}

Figure~\ref{fig:binary-confusion} shows the confusion matrices for binary classification, and Table~\ref{tbl:results} shows the model's performance on all the evaluation metrics described in Section 2.7. Both Figure~\ref{fig:binary-confusion} and Table~\ref{tbl:results} were created with data from five independent training, validation, and test runs of the classifier. Unless otherwise noted, the default threshold for binary classification is 0.5, where classifier confidence equal to or exceeding 0.5 counts as an SNIa classification, and vice versa. Altering this threshold produces Figure~\ref{fig:roc}, the ROC curve of one of these test runs.

\begin{table}[t]
    \centering
    \caption{Evaluation metrics for Ia vs. non-Ia classification on cut dataset}
    \label{tbl:results}
    \begin{tabular}{l c c c}
        \hline
        Metric & Training & Validation & Test \\
        \hline
        Accuracy & $99.93 \pm 0.06$\% & $99.71 \pm 0.2$\% & $99.73 \pm 0.26$\%\\
        Purity & $99.93 \pm 0.06$\% & $99.76 \pm 0.25$\% & $99.68 \pm 0.35$\%\\
        Efficiency & $99.93 \pm 0.05$\% & $99.67 \pm 0.23$\% & $99.78 \pm 0.22$\%\\
        AUC & $1.0 \pm 4.1\text{e-}5$ & $0.9991 \pm 1.6\text{e-}3$ & $0.9994 \pm 1\text{e-}3$ \\
        \hline
    \end{tabular}
\end{table}

\begin{figure}
    \figurenum{6}
    \label{fig:grid}
    \includegraphics[scale=0.5]{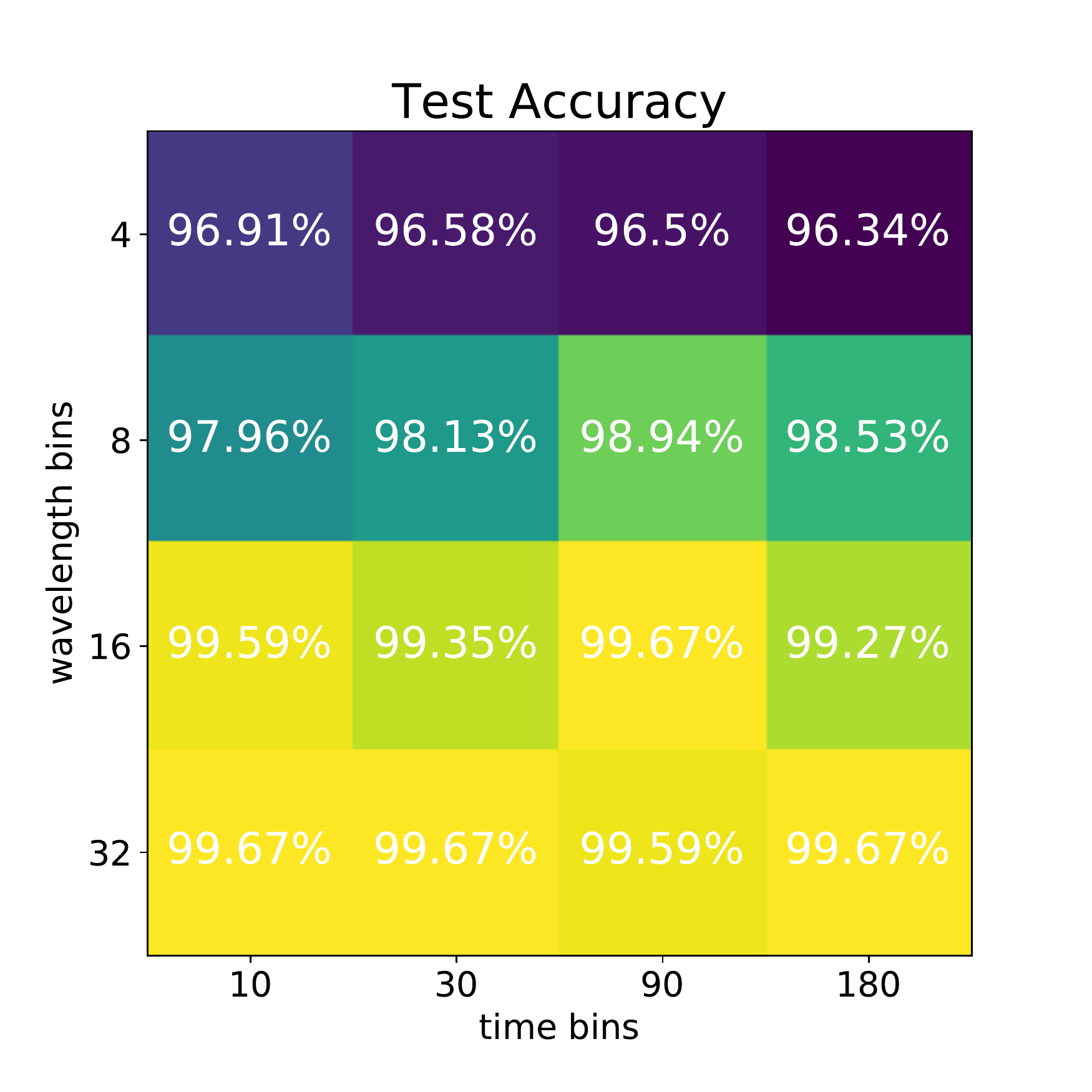}
    \centering
    \caption{Test set accuracies for each choice of wavelength and time bins.}
\end{figure}

\subsubsection{Heatmap Dimensions}
We explored the binary classification performance of different heatmap dimensions in both the time (width) and wavelength (height) axes. For our 7100~\AA\ wavelength range ($3000$~\AA $< \lambda < 10,100$~\AA), we chose intervals of 221.875~\AA, 443.8~\AA, 887.5~\AA, and 1775~\AA, resulting in heatmaps with 32, 16, 8, and 4 wavelength "bins", respectively. For our 180 day range ($t_{\rm peak} - 50 \leq t \leq t_{\rm peak} + 130$ ), we chose intervals of 1, 2, 6, and 18 days, resulting in heatmaps with 180, 90, 30, and 10 time bins.

Figure~\ref{fig:grid} shows the training, validation, and test accuracies for each choice of wavelength and time dimensions. Our classifier seems relatively robust to these changes, showing minimal performance impacts for wavelength bins $\geq 16$ and impressive performance for the smaller sizes as well. Test accuracy drops a maximum of 2.67\% between the best- and worst-performing variants, $32 \times 180$ and $4 \times 10$. This is noteworthy as the $32 \times 180$ heatmaps contain 144 times the number of pixels of the $4 \times 10$ heatmaps.

The performance seems to unilaterally improve as expected as the number of wavelength bins increase, but increasing the number of time bins seems to yield varying, though likely not statistically significant, results.

We have reported all of our \scone\ results using one of the best performing variants, the $32 \times 180$ heatmaps.

\subsubsection{Out-of-Distribution Results}
Preliminary exploration into the out-of-distribution task of training on the spectroscopic dataset and testing on the main dataset  yielded 80.6\% test accuracy. The full results of training on 85\% of the spectroscopic dataset, validating on the remaining 15\%, and testing on the full main dataset are shown in Table \ref{tbl:ood}. Since the redshift distribution of the spectroscopic dataset is skewed toward lower redshifts, testing on class-balanced low-redshift subsets of the main dataset yielded 83\% test accuracy for $z < 0.4$ and 87\% test accuracy for $z < 0.3$.  \cite{Boone_2019} introduced redshift augmentation to mitigate this effect.  The mismatch between the test and training sets, however, comes in other forms.  For example, if the training data is generated using models rather than real data, differences in the characteristics of the spectral surfaces can have an impact on classification.  The unknown relative rates of of each type of event also affect the overall performance.  Since out-of-distribution robustness is an integral part of the challenge of photometric SNe classification, improving the performance of \scone\ on these metrics will be the topic of a future paper.\\\\

\begin{table}[h]
    \centering
    \caption{Evaluation metrics for out-of-distribution Ia vs. non-Ia classification}
    \label{tbl:ood}
    \begin{tabular}{l c c c}
        \hline
        Metric & Training & Validation & Test \\
        \hline
        Accuracy & $99.65 \pm 0.36$\% & $98.84 \pm 1.1$\% & $80.61 \pm 1.75$\%\\
        Purity & $99.86 \pm 0.31$\% & $98.74 \pm 1.31$\% & $81.86 \pm 2.22$\%\\
        Efficiency & $99.67 \pm 0.46$\% & $98.75 \pm 1.27$\% & $80.06 \pm 1.31$\%\\
        AUC & $1.0 \pm 8.9\text{e-}5$ & $0.9939 \pm 9.5\text{e-}3$ & $0.8552 \pm 1.4\text{e-}2$ \\
        \hline
    \end{tabular}
\end{table}

\subsection{Categorical Classification}

In addition to binary classification, \scone\  is able to perform categorical classification and discriminate between different types of SNe. We performed 6-way categorical classification with the same PLAsTiCC dataset used for binary classification as well as the class-balanced dataset described in Section 2.1. Our model differentiated between SN types Ia, II, Ibc, Iax, SN-91bg, and SLSN-1. On the PLAsTiCC dataset (not class-balanced), it achieved $99.26 \pm 0.18$\% training accuracy, $99.13 \pm 0.34$\% validation accuracy, and $99.18 \pm 0.18$\% test accuracy. The confusion matrices in Figure~\ref{fig:categorical-plasticc-confusion} show the average by-type breakdown for five independent runs.

\begin{figure}
    \figurenum{7}
    \label{fig:categorical-confusion}
    \includegraphics[scale=0.35]{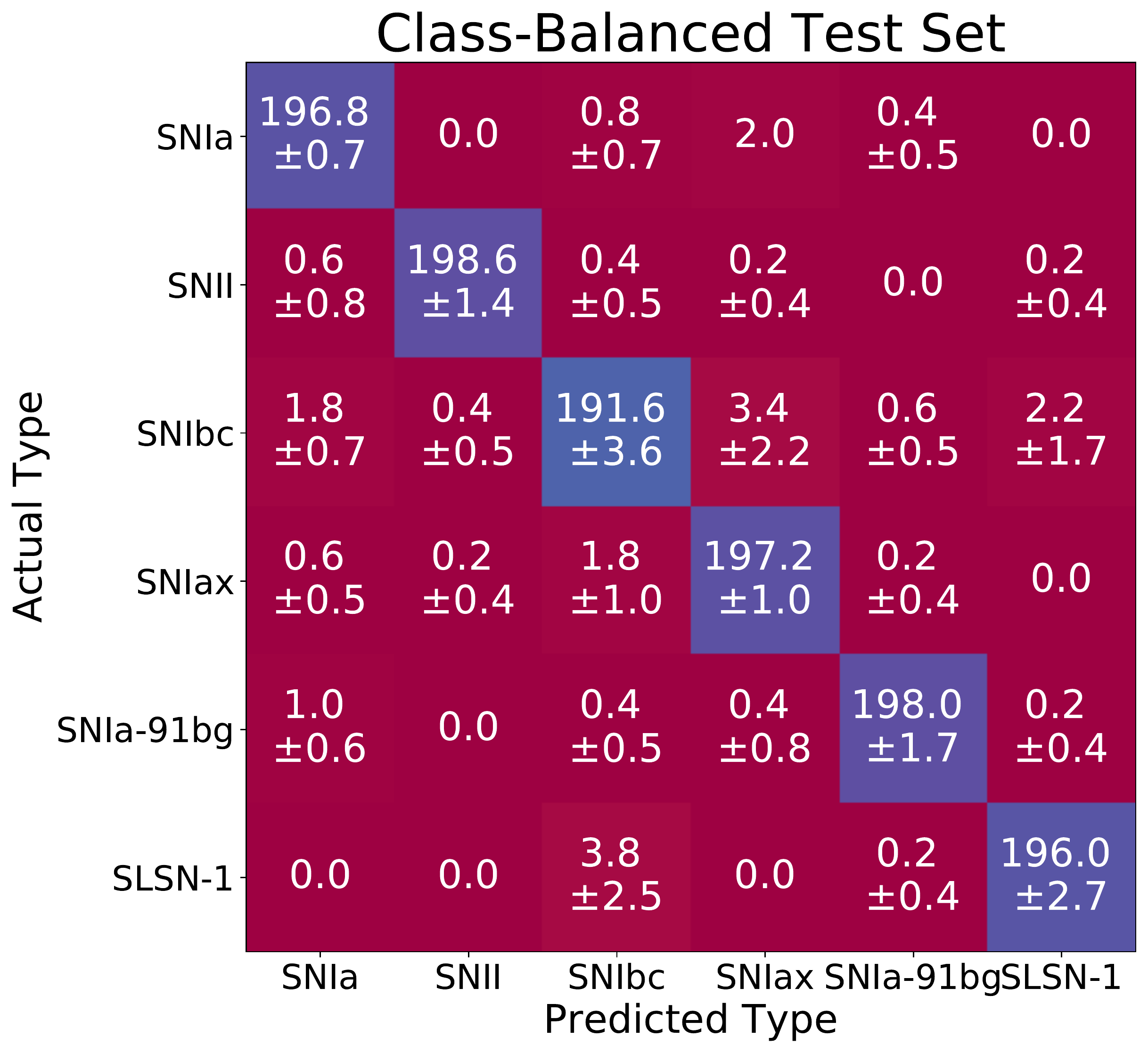}
    \centering
    \caption{Confusion matrix showing average and standard deviation over five runs for categorical classification on the balanced test set.}
\end{figure}

\begin{figure}
    \figurenum{8}
    \label{fig:categorical-plasticc-confusion}
    \includegraphics[scale=0.35]{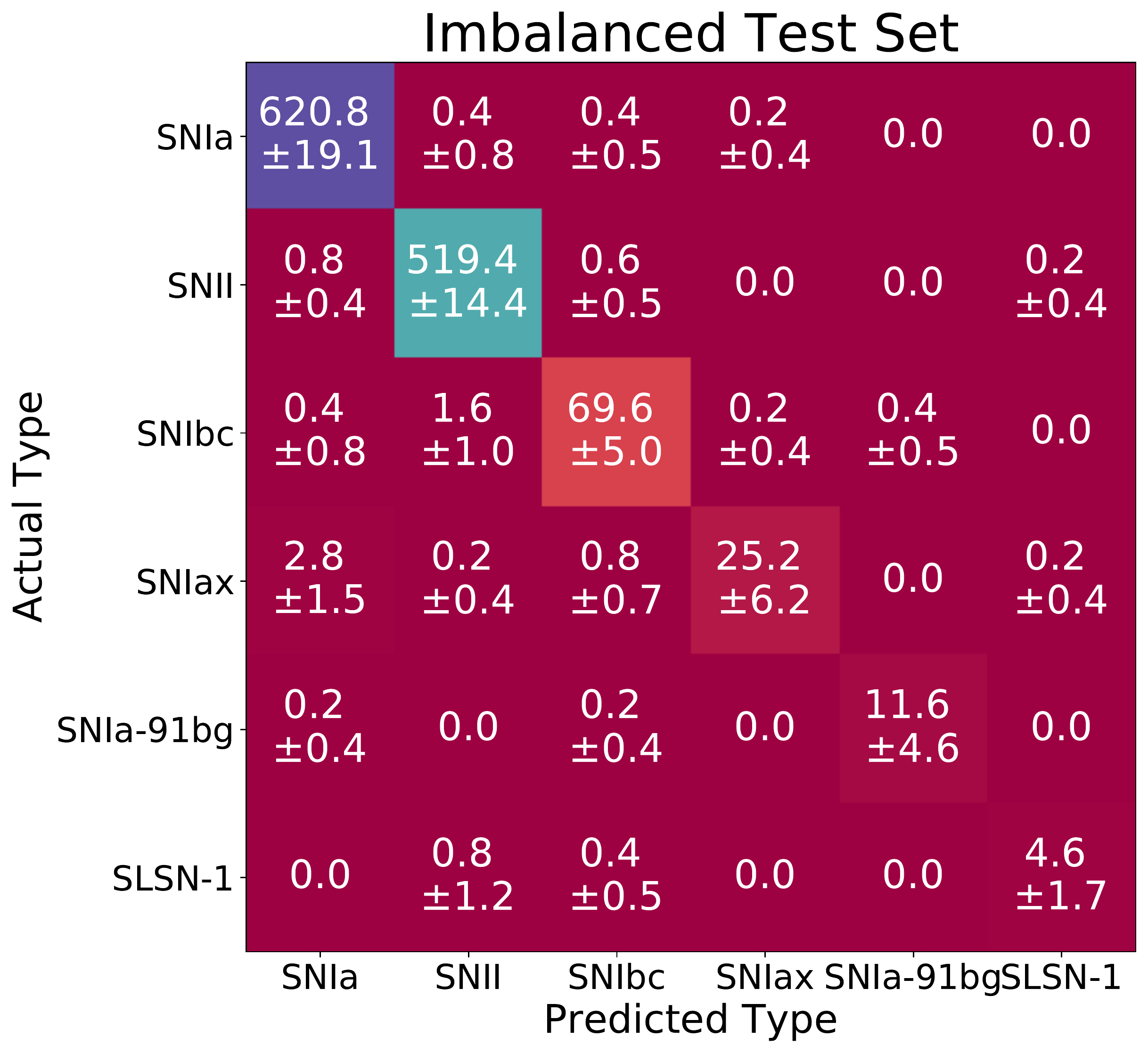}
    \centering
    \caption{Confusion matrix showing average and standard deviation over five runs for categorical classification on the PLAsTiCC (imbalanced) test set.}
\end{figure}

On the balanced dataset, it achieved $97.8 \pm 0.32$\% training accuracy, $98.52 \pm 0.28$\% validation accuracy, and $98.18 \pm 0.3$\% test accuracy. The confusion matrices in Figure~\ref{fig:categorical-confusion} show the average and standard deviations of the by-type breakdown for 5 independent runs.

It is worth noting that we trained and tested with the PLAsTiCC dataset even though it is not class-balanced for this task to try to evaluate the model's performance on a dataset emulating the relative frequencies of these events in nature.

\section{discussion}
Analysis of misclassified heatmaps was performed for both binary and class-balanced categorical classification. No clear evidence of the effect of redshift on accuracy was found for either mode of classification. The quantity of misclassified SNIa examples per run is not sufficient for us to draw conclusions about the accuracy evolution as a function of redshift.

\subsection{Binary Classification}
According to the data presented in Figure~\ref{fig:binary-confusion}, the model seems to mispredict about the same number of Ia's as non-Ia's. An average of 3.33 Ia's are mispredicted in the training set compared with 3.5 non-Ia's. For the validation set, 2 Ia's are mispredicted compared to 1.5, and 1.33 Ia's and 2 non-Ia's are mislabeled for the test set. 
\begin{table*}
    \centering
    \caption{Misclassified test set heatmaps by true and predicted type for binary classification.}
    \label{tbl:misclassified-binary}
    \begin{tabular}{l l c c c c c}
        \hline
        True Type & Predicted Type & $>$90\% Confidence & 90-70\% Confidence & 70-50\% Confidence  & Total & Percentage \\
        \hline
        SNIa & non-Ia & 1 & 0 & 0 & 1 & 25\%\\
        \hline
        SNII & \multirow{2}{0em}{SNIa} & 2 & 0 & 0 & 2 & 50\%\\
        SNIax & & 1 & 0 & 0 & 1 & 25\%\\
        \hline
    \end{tabular}
\end{table*}

\begin{table*}
    \centering
    \caption{Misclassified test set heatmaps by true and predicted type for categorical class-balanced classification.}
    \label{tbl:misclassified-categorical}
    \begin{tabular}{l l c c c c c}
        \hline
        True Type & Predicted Type & $>$90\% Confidence & 90-70\% Confidence & 70-50\% Confidence  & Total & Percentage \\
        \hline
        SNIa & SNIax & 1 & 0 & 0 & 1 & 4.8\%\\
        \hline
        \multirow{2}{0em}{SNII} & SNIa & 1 & 0 & 0 & 1 & 4.8\%\\
        & SNIbc & 0 & 0 & 1 & 1 & 4.8\%\\
        \hline
        \multirow{3}{0em}{SNIbc} & SNIa & 0 & 0 & 1 & 1 & 4.8\%\\
        & SNIax & 1 & 0 & 0 & 1 & 4.8\%\\
        & SLSN-1 & 0 & 0 & 1 & 1 & 4.8\%\\
        \hline
        \multirow{3}{2em}{SNIax}& SNIa & 1 & 2 & 0 & 3 & 14.3\%\\
        & SNIbc & 1 & 2 & 0 & 3 & 14.3\% \\
        & SNIa-91bg & 1 & 0 & 0 & 1 & 4.9\%\\
        \hline
        SLSN-1 & SNIbc & 3 & 1 & 3 & 7 & 33.3\%\\
        \hline
        SNIa-91bg & SNIbc & 0 & 1 & 0 & 1 & 4.8\%\\
        \hline
    \end{tabular}
\end{table*}

The misclassified set summarized in Table~\ref{tbl:misclassified-binary} is the result of one of the five runs represented by the data in Figure~\ref{fig:binary-confusion}. In this example, the model missed 5 examples total during testing -- 1 SNIa, 2 SNII, and 1 SNIax. It was $> 90$\% confident about all of these misclassifications, which is certainly not the case for categorical classification. This could be due to the fact that there are more examples of each type for binary classification than categorical ($\sim 6,000$ and 1,200 per type for training, respectively).


\subsection{Categorical Classification}
The data in Figures~\ref{fig:categorical-confusion} and~\ref{fig:categorical-plasticc-confusion} show some level of symmetry between misclassifications. SNIax and SLSN-1 seem to be easily distinguishable across the board, for example, with 0's in all relevant cells in both figures except one. In Figure~\ref{fig:categorical-confusion}, SNIbc and SLSN-1 are seemingly very similar, as they are misclassified as one another at similarly high rates.

There are notable differences between Figures~\ref{fig:categorical-confusion} and~\ref{fig:categorical-plasticc-confusion}, however. In Figure~\ref{fig:categorical-confusion}, representing the classifier's performance on class-balanced categorical classification, the model mispredicts SNIa's as other types at a similar rate as non-Ia's mispredicted as Ia's. An average of 3.2 Ia's are mispredicted, whereas an average of 4 non-Ia's are misclassified as Ia. In the confusion matrix shown in Figure~\ref{fig:categorical-plasticc-confusion}, significantly more non-Ia's were mispredicted as Ia. An average of 1 Ia was mispredicted compared to an average of 4.2 non-Ia's misclassified as Ia. The rate of SNIbc's mispredicted as SLSN-1 is also significantly lower for the PLAsTiCC dataset than for the balanced dataset. These observations further reinforce the impact of imbalanced classes in classification tasks.

The misclassified set summarized in Table~\ref{tbl:misclassified-categorical} is the result of one of the five class-balanced categorical classification runs represented by the data in Figure~\ref{fig:categorical-confusion}.

One point of interest is the lack of symmetry between misclassifications, in contrast with the analysis of Figures~\ref{fig:categorical-confusion} and~\ref{fig:categorical-plasticc-confusion}. This is clear in the significantly larger number of SLSN-1 misclassified as SNIbc  (7) compared with the number of SNIbc misclassified as SLSN-1 (1). SNIax is also more often misclassified as other types (3 as SNIa, 3 as SNIbc, and 1 as SNIa-91bg) than non-Iax misclassified as Iax (1 SNIa and 1 SNIbc). The more symmetric Figure~\ref{fig:categorical-confusion} suggests that the asymmetry of this table is due to randomness and would be corrected with data from other runs.

The distribution of misclassified examples across the confidence spectrum is non-uniform. In this table, \textit{confidence} refers to the probability assigned to the predicted type by the classifier. Confidence near 100\% for a misclassified example is potentially more insightful than one near 50\%. 9 out of the 21 misclassified heatmaps were misclassified at $> 90$\% confidence, 6 at 90-70\% confidence, and 6 at 70-50\%. Surprisingly, the classifier is confidently wrong almost half the time. One particularly interesting example is a SNIbc "misclassified" as SLSN-1, but the classification probabilities for both SLSN-1 and SNIbc were 50\%.

\subsection{Limitations and Future Work}
As stated in section 2.1, it is important to note that the metrics reported in this paper are in-distribution results since the training, validation, and test sets are mutually exclusive segments of the main dataset. The out-of-distribution performance of \scone, as evaluated in section 3.1.2, is noticeably diminished from the $>99$\% in-distribution test accuracy. The high in-distribution test accuracy shows that \scone\ is robust to previously unseen data, but the lower out-of-distribution test accuracy demonstrates \scone's sensitivity to variations in the parameters of the dataset, such as the redshift distribution, relative rates of different types of SNe, small variations in the SN~Ia model, as well as telescope characteristics. Generalizing \scone\ to become robust to these variations will be the subject of a future paper.

\section{Conclusions}
In this paper we have presented \scone, a novel application of deep learning to the photometric supernova classification problem. We have shown that \scone\  has achieved unprecedented performance on the in-distribution Ia vs. non-Ia classification problem and impressive performance on classifying SNe by type without the need for accurate redshift approximations or handcrafted features.

Using the wavelength-time flux and error heatmaps from the Gaussian process for image recognition also allows the convolutional neural network to learn about the development of the supernova over time in all filter bands simultaneously. This provides the network with far more information than a photograph taken at one moment in time. Our choice of an $h \times 3$ convolutional kernel, where $h$ is the number of wavelength bins, supplements these benefits by allowing the network to learn from data on the full spectrum of wavelengths in a sliding window of 3 days.

As future large-scale sky surveys continue to add to our ever-expanding transients library, we will need an accurate and computationally inexpensive photometric classification algorithm. Such a model can inform the best choice for allocation of our limited spectroscopic resources as well as allow researchers to further cosmological science using minimally contaminated SNIa datasets. \scone\  can be trained on tens of thousands of lightcurves in minutes and confidently classify thousands of lightcurves every second at $> 99$\% accuracy.

Although \scone\  was formulated with supernovae in mind, it can easily be applied to classification problems with other transient sources. The documented source code has been released on Github (github.com/helenqu/scone) to ensure reproducibility and encourage the discovery of new applications.

\acknowledgments
The authors would like to thank Rick Kessler for his help with SNANA simulations and Michael Xie for his guidance on the model architecture. This research used resources of the National Energy Research Scientific Computing Center (NERSC), a U.S. Department of Energy Office of Science User Facility located at Lawrence Berkeley National Laboratory, operated under Contract No. DE-AC02-05CH11231. This work was supported by DOE grant DE-FOA-0001781 and NASA grant NNH15ZDA001N-WFIRST.


    
\bibliography{references}
\end{document}